\documentclass[aps,pre,amsmath,amssymb,twocolumn]{revtex4}
\usepackage{xcolor}
\usepackage{graphicx}
\usepackage{amsmath}

\newcommand{\be}{\begin{equation}}
\newcommand{\ee}{\end{equation}}
\newcommand{\ba}{\begin{eqnarray}}
\newcommand{\ea}{\end{eqnarray}}
\newcommand{\baa}{\begin{eqnarray}}
\newcommand{\eaa}{\end{eqnarray}}
\newcommand{\ed}{\end{document}}
\newcommand{\lab}[1]{\label{#1}}
\newcommand{\re}[1]{(\ref{#1})}
\newcommand{\ci}[1]{\cite{#1}}

\begin{document}
\title{PT-symmetric dynamical confinement:\\ Fermi acceleration, quantum force and Berry phase}

\author{S.\ Rakhmanov$^{a}$, C.\ Trunk$^b$, M.\ Znojil$^c$ and D.\ Matrasulov$^{d,f}$}

\thanks{\includegraphics[height=7.0mm]{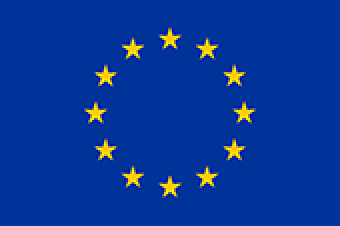} This paper is supported by European Union’s Horizon 2020 research and innovation programme under the Marie Sklodowska-Curie grant agreement ID: 873071, project SOMPATY (Spectral Optimization: From Mathematics to Physics and Advanced Technology).
The work of SR is partially supported by the Grant REP-05032022/235, funded under the MUNIS Project, supported by the World Bank and the Government of the Republic of Uzbekistan.
}

\affiliation{$^a$Chirchik State Pedagogical University, 104 Amur Temur Str., Chirchik, 111700,Uzbekistan\\ $^b$Technische Universit\"at Ilmenau, Postfach 100565, 98684 Ilmenau, Germany  \\ $^c$Nuclear Physics Institute, Czech Academy of Sciences, 25068 \v{R}e\v{z}, Czech Republic \\ $^d$Turin Polytechnic University in Tashkent, 17 Niyazov Str., 100095,  Tashkent, Uzbekistan \\ $^f$Center for Theoretical Physics, Khazar University, 41 Mehseti Street, Baku, AZ1096, Azerbaijan}

\begin{abstract}

We consider a quantum particle under the dynamical confinement caused by PT-symmetric box with a moving wall.
The latter is described in terms of the time-dependent Schr\"{o}dinger equation  obeying the time-dependent PT-symmetric boundary conditions. The class of the functions, describing time-dependence of the wall's position and keeping the system as PT-symmetric is found. Physically observable characteristics, such as average kinetic energy and the average quantum force are calculated as a function of time. Also, geometric phase is calculated for the harmonically  oscillating wall regime. Experimental realization of the proposed model is discussed.
\end{abstract}

\maketitle

\section{Introduction}
Since the publication of
the pioneering paper by Bender and Boettcher  \ci{Bender98},
parity-time symmetric quantum systems attracted much attention
\ci{Bender07} - \ci{[8]}.
It was long believed 
that only Hermitian Schr\"{o}dinger  operators can
describe physical systems. However, 
Bender with Boettcher 
refuted such a statement. 
They argued
that also some non-Hermitian but PT-symmetric operators could have real eigenvalues and
play, therefore, the role of acceptable quantum Hamiltonians.

This conjecture caused rapidly growing interest in PT-symmetric quantum physics. Different aspects of the topic have been studied both in quantum mechanical and field theoretical contexts. A huge number of papers dealing with various issues of PT-symmetric quantum physics have been published during the past two decades. 
These studies allowed one to construct various mathematically consistent theories of PT-symmetric quantum systems including even several innovative non-Hermitian but PT-symmetric quantum field theories \ci{Bender99}-\ci{Ours19_02}. 

Experimental realization of such systems was also the subject of extensive research. The latter has been done mainly in optics \ci{[6],Ganainy07,Ruter10}. Some other PT-symmetric systems have been discussed recently in the literature \ci{Znojil09,Znojil17_03,Znojil17_04,Znojil22_01}. General theory of  PT-symmetric quantum mechanics have been developed introducing so-called PT-symmetric inner product. However, since such a
condition does not provide a positively
definite norm,
its extension in
terms of a PCT-symmetric inner product was proposed in \ci{Bender02_02,Mostafazadeh02,Bender07,Bender08_02,Znojil08}.

Within such a theory a PT-symmetric model 
can be introduced either through the complex potential, or by imposing proper boundary conditions, which provide such symmetry via the inner product. Different types of complex potentials that provide PT-symmetry in the Hamiltonian have been considered in \ci{Bender07},
\ci{Bender08_02}. Introducing PT-symmetry in terms of proper boundary conditions was studied for the particle-in-box systems in \ci{Bender18_01} or \ci{Znojil91, Znojil05,Znojil06,Znojil12}.
 
In this paper we propose a model for dynamical confinement in a PT-symmetric well by considering hard-wall box with moving boundaries. The 
system is assumed to satisfy PT-symmetric hard-wall boundary conditions. The 
corresponding quantum dynamics
is described in terms of time-dependent Schr\"{o}dinger equation 
upon which time-dependent boundary conditions are imposed. In other words, we will
solve the Schr\"{o}dinger equation endowed
with 
PT-symmetric, time-dependent boundary conditions.

\section{Hermitian model}

We note that the conventional Hermitian Schr\"{o}dinger equation with time-dependent boundary conditions have been the topic for extensive research during past four decades, especially, for 1D case (see, e.g., Refs.\ci{doescher}-\ci{Uzy22}). 
Early treatments of the dynamical confinement in nonrelativistic quantum mechanics date back to Doescher, who studied basic aspects of the problem \cite{doescher}. Munier et al.\ studied  more detailed research of the problem and calculated physically observable quantities for the problem of time-dependent box \cite{mun81}. Later, Makowsky \cite{mak91}-\cite{mak923} and Razavy \cite{razavy01,razavy02} provided  a systematic study of the dynamical confinement problem, by considering one-dimensional box with moving walls and classifying time-dependence of the wall approving exact solution of the Schr\"{o}dinger equation with time-varying boundary conditions.  Unitary transformation that maps the time-dependent box to that with fixed walls, was found in \cite{razavy01,razavy02}. These studies used an approach developed earlier by Berry and Klein \cite{berry00}. Different aspects of the problem dynamical confinement and its applications to dynamical Casimir effect was studied in a series of papers by Dodonov et al.\ \cite{dodonov01}-\cite{dodonov04}. Berry phase in time-dependent box was studied in \cite{per,Kwon,Wang}. \v Seba studied the problem of time-dependent box in the context of quantum Fermi acceleration \cite{seb90}. The quantum gas under the dynamical confinement was considered in \cite{nakamura01,nakamura02}, where quantum force operator for time-dependent box was introduced. 
Hydrogen-like atom confined in a spherical hard-wall box with time-varying radius was considered in \cite{Our03}. Different aspects of the problem of dynamical confinement were studied also in \cite{pinder} -\cite{glas09}.  The problem of time-dependent Neumann boundary conditions is considered in \ci{duca}. Extension of the dynamical confinement to relativistic case   by considering Dirac equation for time-dependent box  was done in \cite{sobirov}. Time-dependent quantum graphs have been considered in the  Refs. \cite{matrasulov,nikiforov,Uzy22}.


Let us now
briefly recall Hermitian counterpart of our model the one-dimensional problem of studying the quantum dynamics of a particle confined in box with time-varying wall, $L(t)$ following the Ref. \ci{mak91}. The dynamics of the particle in such system can be described by the time-dependent Schr\"odinger equation ($\hbar=m=1$)
\begin{equation}
    i\frac{\partial\Psi(x,t)}{\partial t}=-\frac12\frac{\partial^2\Psi(x,t)}{\partial x^2}, \quad t\geq 0, x \in {\mathbb R} \label{shro01}
\end{equation}
with time dependent Dirichlet boundary conditions
\begin{equation}
\Psi(x,t)|_{x=0}=\Psi(x,t)|_{x=L(t)}=0,\, \text{\ for all \ }  t\in {\mathbb R}^+. 
\end{equation}
Here, $\Psi(x,t)$ is the wave function of the quantum particle.
Introducing a new coordinate, $y=\frac{x}{L(t)}$ allows us to get the wave equation imposed static boundary conditions as
\begin{equation}
    i\frac{\partial\Psi(y,t)}{\partial t}=-\frac{1}{2L^2}\frac{\partial^2\Psi(y,t)}{\partial y^2}+i\frac{\dot{L}}{L}y\frac{\partial\Psi(y,t)}{\partial y} \label{shro02}   
\end{equation}
with new boundary conditions given by
\begin{equation}
\Psi(y,t)|_{y=0}=\Psi(y,t)|_{y=1}=0,\, \text{\ for all \ }  t\in {\mathbb R}^+ .    
\label{bc02}
\end{equation}

The Hamilton operator in the Eq.\ (\ref{shro02}) is not self-adjoint. To rebuild the self-adjointness one uses the transformation 
\begin{equation}
\Psi(y,t)=\sqrt{2/L}\exp{\left(\frac{i}{2}L\dot{L}y^2\right)}\varphi(y,t),
\end{equation}
where $\dot{L}=dL/dt$. Then we get \ci{mak91}
\begin{equation}
    i\frac{\partial\varphi}{\partial t}=-\frac{1}{2L^2}\frac{\partial^2\varphi}{\partial y^2}+\frac12L\ddot{L}y^2\varphi ,   
\end{equation}
where $\ddot{L}=d\dot{L}/dt$ and $\varphi(y,t)$ satisfies the boundary conditions (\ref{bc02}).

\section{Static PT-symmetric  quantum box}
\label{Carsten}



Time-independent counterpart of our system, i.e.\ the PT-symmetric particle-in-box system has been  considered earlier in the Refs. \ci{Znojil06,Znojil12,Dasarathy13,Znojil15}. Here we will briefly recall its description following the Ref. \ci{Znojil12}.
For one-dimensional Schr\"{o}dinger equation given as
\begin{equation}
    -\frac{1}{2}\frac{d^2\Psi(x)}{d x^2}=E\Psi(x) \label{seq1}.
\end{equation}

PT-symmetry of the system is provided by Robin type boundary condition which is given as \ci{Znojil12}

\begin{equation}
\begin{split}
   & \left.\left(\frac{d\Psi(x)}{d x}+i\alpha\Psi(x)\right) \right|_{x=-L}\\
   &=\left.\left(\frac{d\Psi(x)}{d x}+i\alpha\Psi(x)\right)\right|_{x=L}=0,    
   \label{srbc01}
    \end{split}
\end{equation}
and $\alpha>0$.

The solution of Eq.\ \re{seq1} fulfilling the PT-symmetric boundary condition given by Eq.\ \re{srbc01} can be written as

\begin{equation}
    \Psi_n(x)=A_n\left( \sin\frac{\pi n x}{L} +\frac{i\pi n}{L\alpha}\cos\frac{\pi n x}{L} \right), \label{ssol}
\end{equation}
here the $A_n$ is the normalization constant given by
$$
 A_n=\sqrt{\frac{L\alpha^2}{L^2\alpha^2+\pi^2n^2}}.
$$
Corresponding eigenvalues can be written as
\begin{equation}
    E_n=\frac{\pi^2n^2}{2L^2}, \label{eig}
\end{equation} 
where $n$ is the quantum number.

\section{PT-symmetric quantum box with a moving wall}


A time-dependent version of the above PT-symmetric boundary condition given by Eq.\ \re{srbc01}  can be written as
\begin{equation}
\begin{split}
& \left.\left(\frac{\partial\Psi(x,t)}{\partial x}+i\alpha\Psi(x,t)\right)\right|_{x=-L(t)}\\
&=\left.\left(\frac{\partial\Psi(x,t)}{\partial x}+i\alpha\Psi(x,t)\right)\right|_{x=L(t)}=0
\label{rbc01}
\end{split}
\end{equation}
for some  $\alpha>0$ and for all $t\in{\mathbb R}^+$, where the time evolution is given in terms of the Schr\"odinger equation \eqref{shro01}.
Following the approach from Section~\ref{Carsten}, one has here 
a kind of $PT$-symmetry. Namely let
$P$ be the parity and $T$ the time reversal operators, which are defined as
\begin{equation}
    \begin{split} 
    & P\Psi(x,t)=\Psi(-x,t), \\
    & T\Psi(x,t)=\Psi^*(x,-t). 
    \end{split}
\end{equation}
If
\begin{equation}\label{Carsten3}
H:=-\frac{1}{2}\frac{\partial^2}{\partial x^2}
\end{equation}
denotes the Hamilton operator.

Here $PT$-symmetry means that
whenever $\Psi(x,t)$ is a wave function satisfying 
the above $PT$-symmetric boundary condition \eqref{rbc01}, then 
one has
\begin{equation} \label{Carsten2}
    PTH\Psi(x,t)=HPT\Psi(x,t)
\end{equation}
and the function $PTH\Psi(x,t)$ also satisfies
the $PT$-symmetric boundary condition \eqref{rbc01}.

We show that this is satisfied for the Hamiltonian in \eqref{Carsten3}
and $PT$-symmetric boundary condition \eqref{rbc01}.
Indeed, using the definition of $P$ and $T$, it is easy to see
\begin{equation*}
    PTH\Psi(x,t)=HPT\Psi(x,t)=-\frac{1}{2}\frac{\partial^2}{\partial x^2}\Psi^*(-x,-t),
\end{equation*}
which shows \eqref{Carsten2}. Moreover, if a wave function $\Psi(x,t)$ satisfies 
the boundary condition \eqref{rbc01}, then
\begin{equation*}
\begin{split}
&PT\left( \frac{\partial \Psi(x,t)}{\partial x}+i\alpha\Psi(x,t)\right)\left|^{x=-L(t)}_{x=L(t)}\right.\\[1ex]
&=\left( -\frac{\partial \Psi^*(-x,-t)}{\partial x}+i\alpha\Psi^*(-x,-t)\right)\left|^{x=-L(t)}_{x=L(t)}\right.\\[1ex]
&     =-\left( \frac{\partial \Psi(-x,-t)}{\partial x}+i\alpha\Psi(-x,-t)\right)^*\left|^{x=-L(t)}_{x=L(t)}\right. .
\end{split}
\end{equation*}
One can see from equation above, this system is $PT$-symmetric, if $L(t)$ is an even function, i.e.\ $L(t)=L(-t)$,
\begin{equation}
    -\left(\Psi'(\pm L(-t),-t)+i\alpha\Psi(\pm L(-t),-t)\right)^*=0 . 
\end{equation}

We note that norm conservation is broken for the boundary conditions \re{rbc01}:
\begin{eqnarray*}
    \begin{split}
     &\frac{\partial N}{\partial t}=\frac{\partial}{\partial t}\int_{-L(t)}^{L(t)}|\Psi(x,t)|^2dx=\int_{-L(t)}^{L(t)}
\frac{\partial}{\partial t}|\Psi(x,t)|^2dx\\[1ex]
&     +
 \dot{L}(t)\left( |\Psi(L(t),t)|^2+|\Psi(-L(t),t)|^2 \right)\\[1ex]
 &   =(\dot{L}(t)+\alpha) |\Psi(L(t),t)|^2+(\dot{L}(t)-\alpha)|\Psi(-L(t),t)|^2,
    \end{split}
    \end{eqnarray*}
which is non-zero, hence the norm is not conserved.

Solving the Eq.\ (\ref{shro01}) with the Robin-type boundary conditions (\ref{rbc01}) is difficult because of the time-dependent boundary conditions. In order to change the time-dependent boundary conditions into static ones we use two transformations. One of them is for the coordinate and  is given by
\begin{equation}
    y=\frac{x}{L(t)}.  \label{trans01}  
\end{equation}
Substituting the transformation in  (\ref{trans01}) into the equation gives 
\begin{equation}
    i\frac{\partial\Psi(y,t)}{\partial t}=-\frac{1}{2L^2}\frac{\partial^2\Psi(y,t)}{\partial y^2}+i\frac{\dot{L}}{L}y\frac{\partial\Psi(y,t)}{\partial y} 
    \lab{se01}
\end{equation}
with 
\begin{equation}
\begin{split}
    &\left.\left(\frac{1}{L}\frac{\partial\Psi(y,t)}{\partial y}+i\alpha\Psi(y,t)\right)\right|_{y=-1} \\
    &=\left.\left(\frac{1}{L}\frac{\partial\Psi(y,t)}{\partial y}+i\alpha\Psi(y,t)\right)\right|_{y=1}=0 
    \end{split}
    \lab{bcs01}
\end{equation}
for all $t\in{\mathbb R}^+$ and $\alpha>0$. 

Using the second transformation for the wave function 
\begin{equation}
    \Psi(y,t)=e^{-i\alpha L(t) y}\psi(y,t)
\end{equation}
We get
\begin{equation}
    i\frac{\partial\psi}{\partial t}=-\frac{1}{2L^2}\frac{\partial^2\psi}{\partial y^2}+i\frac{\alpha+\dot{L}y}{L}\frac{\partial\psi}{\partial y} +\frac{1}{2}\alpha^2\psi   \label{shro05} 
\end{equation}
with static Neumann boundary conditions as
\begin{equation}
\left.\frac{\partial\psi(y,t)}{\partial y}\right|_{y=-1}=\left.\frac{\partial\psi(y,t)}{\partial y}\right|_{y=1}=0    
\end{equation}
for all $t\in{\mathbb R}^+$.

Eq.\ (\ref{shro05}) can be solved by expanding the wave function in terms of the static box eigenfunctions $\varphi_n(y)=\cos \pi ny$ fulfilling the Neumann boundary conditions $\frac{d\varphi}{dy}|_{y=-1}=\frac{d\varphi}{dy}|_{y=1}=0$
\begin{equation}
    \psi(y,t)=\sum_nC_n(t)\varphi_n(y). \label{expand}
\end{equation} 
By substituting (\ref{expand}) into Eq.\ (\ref{shro05}), we obtain
\begin{equation}
\begin{split}
    & iL^2\sum_n\dot{C}_n(t)\varphi_n(y)=\sum_nC_n(t)\left(-\frac{1}{2}\frac{\partial^2 \varphi_n(y)}{\partial y^2} \right)\\
    & +iL(\alpha+\dot{L}y)\sum_nC_n(t)\frac{\partial\varphi_n(y)}{\partial y}+\frac{1}{2}\alpha^2L^2\sum_nC_n(t)\varphi_n(y).
    \end{split}
\end{equation}

\begin{figure}[t!]
\centering
\includegraphics[totalheight=0.35\textheight]{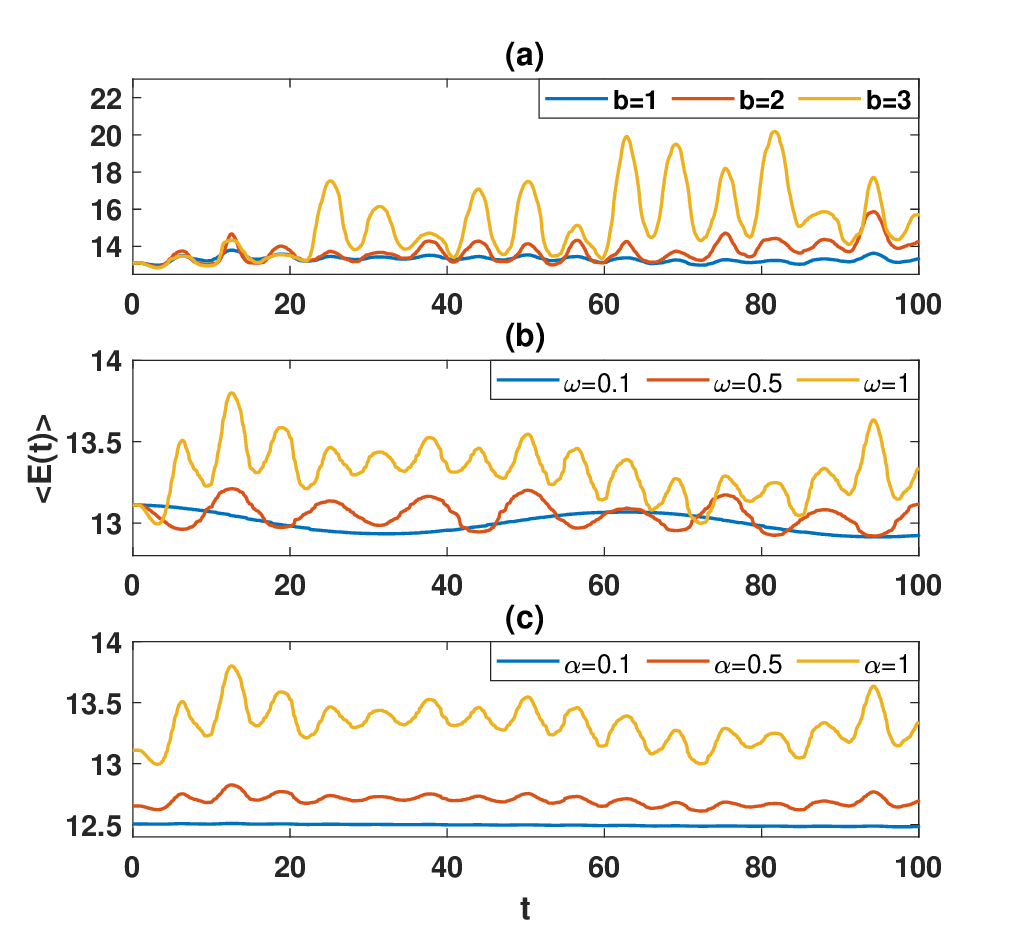}
 \caption{The time dependence of average energy for harmonic oscillating box for different values of oscillating amplitude for $L=10+b\cos t$, $\alpha=1$ (a), frequency for $L=10+\cos\omega t$, $\alpha=1$ (b) and parameter $\alpha$ for $L=10+\cos t$ (c)} \label{fig1}
\end{figure}

Multiplying both sides of this equation to $\varphi_m^*(y)$, integrating over $y$ and using the orthonormal property, $\int_{-1}^1\varphi_m^*(y)\varphi_n(y)dy=\delta_{mn}$, 
we obtain asystem of first order differential equations
for the expansion coefficients $C_n(t)$,
\begin{equation}
     iL^2\dot{C}_n(t)=\frac{\pi^2 n^2+\alpha^2L^2}{2}C_n(t)+\sum_mV_{nm}C_m(t),
     \lab{system1}
\end{equation}
where
\begin{equation}
    V_{nm}=-iL\pi m(\alpha I_{1nm}+\dot{L}I_{2nm})
\end{equation}
with
\begin{equation}
    I_{1nm}=\int_{-1}^1\cos(\pi ny)\sin(\pi my)dy=0
\end{equation}
and
\begin{equation}
\begin{split}
    & I_{2nm}=\int_{-1}^1y\cos(\pi ny)\sin(\pi my)dy\\
    & =\left\{\begin{split}
       &-\frac{1}{2\pi n},  \;\;\;\; n=m, \\
       &-\frac{(-1)^{m+n}}{\pi(m+n)}-\frac{(-1)^{m-n}}{\pi(m-n)}, \;\;\;\; n\neq m\,.
    \end{split}\right.
    \end{split}
\end{equation}

\begin{figure}[t!]
\centering
\includegraphics[totalheight=0.35\textheight]{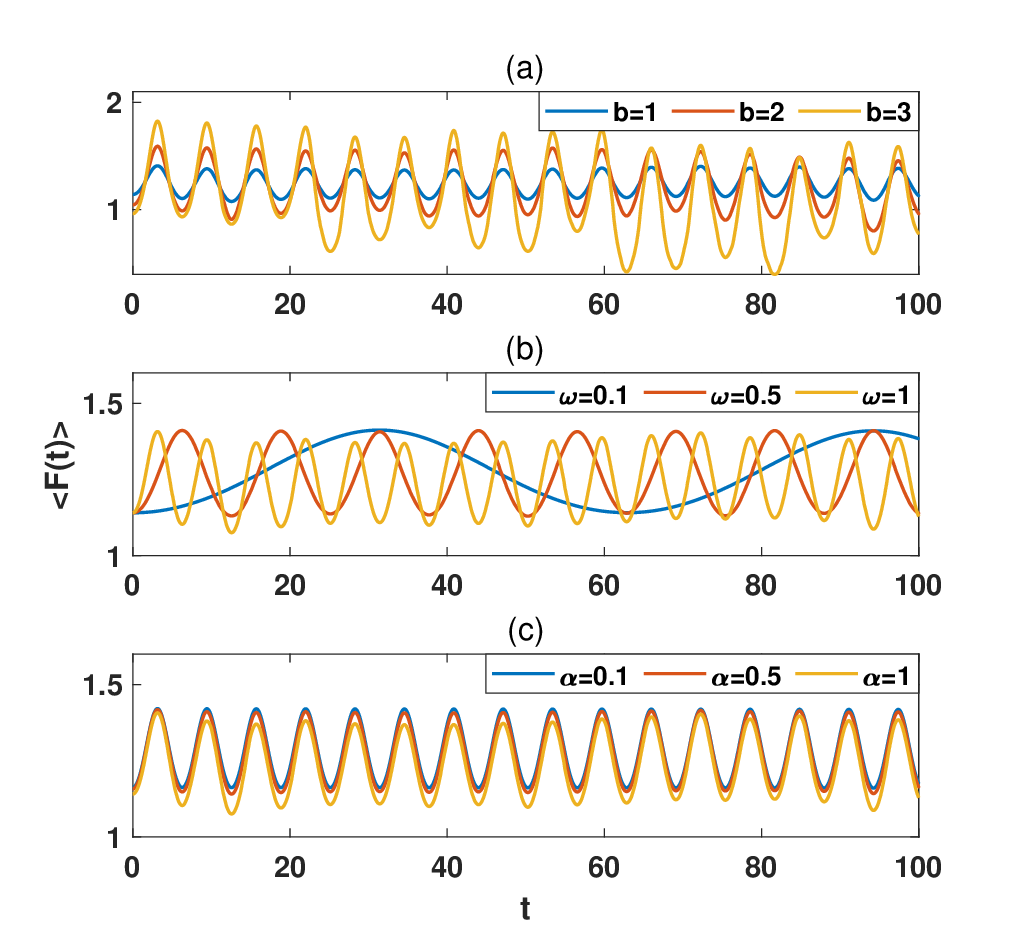}
 \caption{The time dependence of average force for harmonic oscillating box for different values of oscillating amplitude for $L=10+b\cos t$, $\alpha=1$ (a), frequency for $L=10+\cos\omega t$, $\alpha=1$ (b) and parameter $\alpha$ for $L=10+\cos t$ (c)} \label{fig4}
\end{figure}

\section{Quantum dynamics under PT-symmetric time-dependent confinement}

Quantum dynamics of a particle under PT-symmetric confinement can be studied by analyzing time-dependence of such characteristics as the average kinetic energy, average quantum force and (for adiabatically evolving system), so-called geometric phase, or Berry phase.
Average kinetic energy describes acceleration or deceleration of the particle under the dynamical confinement. In case of harmonically oscillating wall, it serves as the main characteristic of so-called Fermi acceleration. The average quantum force describes the expectation value of the force acting by moving wall of the box to a particle confined inside the box. Geometric phase appears in a quantum system, where the evolution is caused by a slowly varying parameter.

\subsection{The average kinetic energy}

The average kinetic energy is determined as the expectation value of the kinetic energy operator:
\be 
<E(t)> = \langle \Psi| H | \Psi \rangle, 
\ee
where $H$ is as in \eqref{Carsten3}.

For PT-symmetric quantum box with a moving wall, using Eqs. \re{expand} and \re{system1}, the average kinetic energy can be written as
\begin{widetext}
$$<E(t)>=\int_{-L(t)}^{L(t)}\Psi^*H\Psi dx=\int_{-1}^{1}e^{i\alpha L(t)y}\psi^*\hat{H}e^{-i\alpha L(t)y}\psi L(t)dy =$$
$$\frac{1}{L(t)}\int_{-1}^{1}e^{i\alpha L(t)y}\sum_nC^*_n(t)\varphi^*_n\left(-\frac{1}{2}\frac{\partial^2}{\partial y^2}\right)e^{-i\alpha L(t)y}\sum_mC_m(t)\varphi_mdy=$$
\begin{equation}
     \frac{1}{L(t)}\sum_{n,m}C^*_n(t)C_m(t)\int_{-1}^{1}\varphi^*_n\left(\frac{\pi^2m^2+\alpha^2L(t)^2}{2}\right)\varphi_mdy=\frac{1}{2L(t)}\sum_n\left(\pi^2n^2+\alpha^2L(t)^2\right)|C_n(t)|^2.
\lab{average}
\end{equation}
\end{widetext}

It is clear that $<E(t)>$ is a real function of time, despite the complex boundary conditions in Eq.\ \re{rbc01}. Also, it is easy to see that for $\alpha =0$ the expression for $<E(t)>$ coincides with that for its Hermitian counterpart (see, e.g.\ Refs. \ci{doescher,nakamura01}).

The average kinetic energy as a function of time is presented in Fig. 1 for (a) different values of the amplitude, (b) frequency and {c} PT-symmetric para
meter, $\alpha$. For all cases, $<E(t)>$ fluctuates in time and there is  no monotonic growth or decay of the average kinetic energy as a function of time. This is a specific property of PT-symmetric systems: Pumping of energy does not cause monotonic growth, as the system undergoes to losses.

\begin{figure}[t!]
\centering
\includegraphics[totalheight=0.35\textheight]{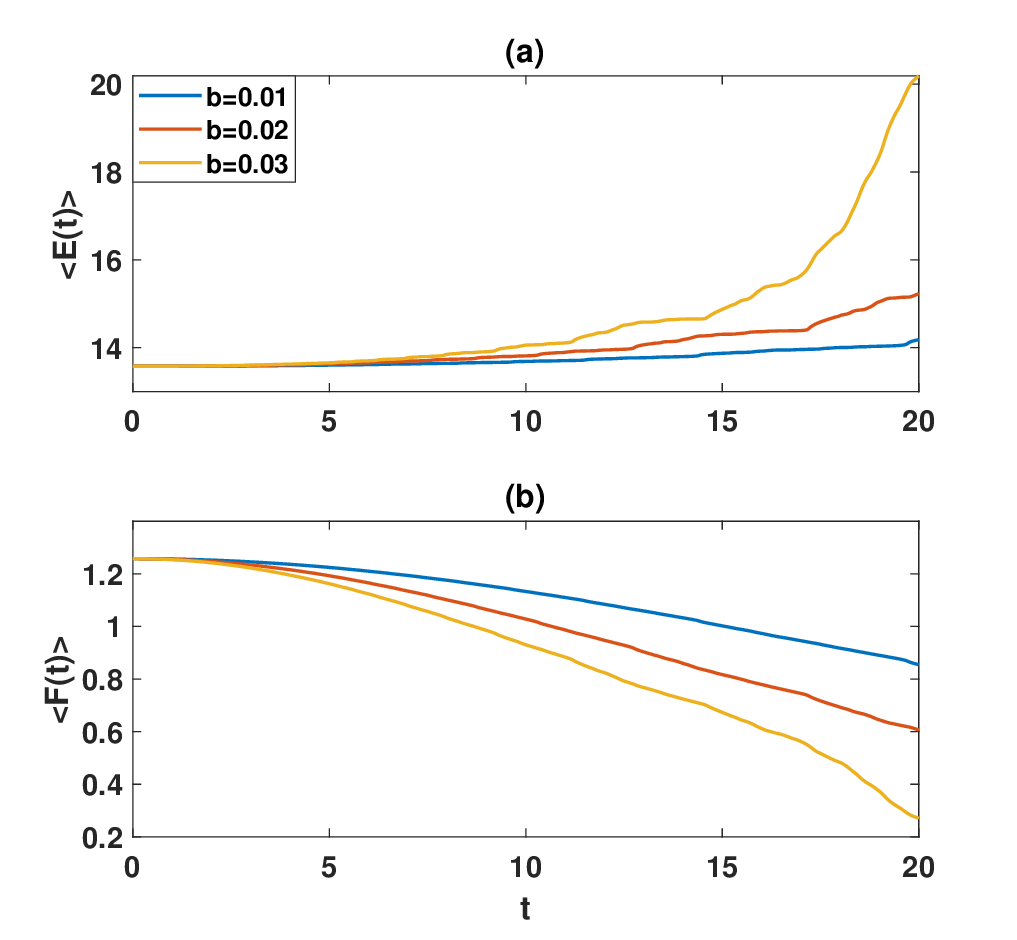}
 \caption{The time dependence of average energy (a) and force (b) for expanding box for different values of $b$ for $L=10+bt^2$, $\alpha=1$} \label{fig2}
\end{figure}

\begin{figure}[t!]
\centering
\includegraphics[totalheight=0.35\textheight]{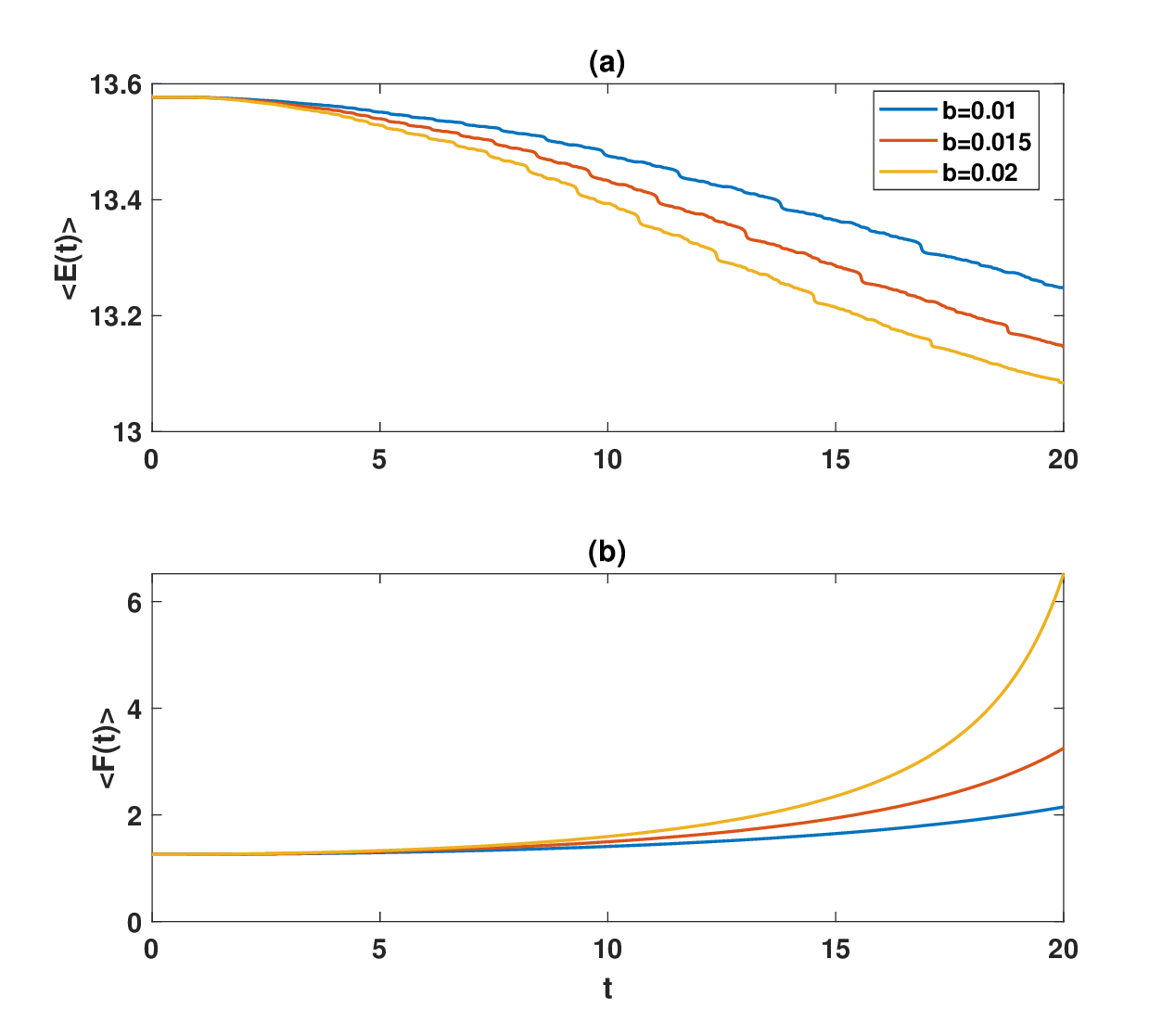}
 \caption{The time dependence of average energy (a) and force (b) for contracting box for different values of $b$ for $L=10-bt^2$, $\alpha=1$} \label{fig3}
\end{figure}

\subsection{The average force}

During its motion the wall of the box interacts with confined particle. This interaction appears in the form of pressure by the wall to the particle. Such pressure is caused by the quantum force which is described in terms of the force operator as
\be 
F=-\frac{\partial  H}{\partial L(t)}   . \lab{for}
\ee

The expectation value of the force operator can be written as \ci{nakamura01}
\begin{equation}
    <F(t)>=-\frac{\partial \langle E(t) \rangle}{\partial L}=\frac{1}{2}\sum_n\left(\frac{\pi^2n^2}{L(t)^2}-\alpha^2\right)|C_n(t)|^2.
    \label{force}
\end{equation}

In Fig.\ 2  plots of $<F(t)>$ are presented as a function of time for harmonically oscillating wall at different values of the oscillation amplitude (a), oscillation frequency (b) and parameter $\alpha$ (c). Unlike the behavior of $<E(t)>$ the time-dependence of
$<F(t)>$ demonstrates certain quasi-periodicity in (a) and (b). Even periodic time-dependence can we observed in the plots of $<F(t)>$ in Fig.\ 2c.

An important case for the above model is monotonically expanding or contracting which is directly related, e.g.\ to the problem of the behaviour of matter under the very heavy pressure and cooling of quantum ideal gas. Fig.\ 3 presents the plots of the average kinetic energy (a) and force (b) as a function of time for quadratically (in time) expanding box $L=a+bt^2$ at different values of acceleration of the wall $b$. Unlike the case of the Hermitian counterpart, where contracting of the box's size causes growth of the average kinetic energy, Fig.\ 3a demonstrates decrease of $<E(t)>$.  This happens due to the second term in Eq.\ \re{average}, which contains the factor $L^(t)$,  growing much faster than $C_n(t)$.

Completely opposite behaviour can be seen in time-dependence of $<F(t)>$ presented in Fig.\ 3b. Here the average force decays in time like in case of Hermitian counterpart of our system. This can be explained by the fact that second term in Eq.\ \re{force} dos not depend on $L(t)$, while first contains the  factor $L^2(t)$, which growth much faster than that of the second term. This can be clearly seen from Fig.\ 5a which compares $L(t)$ and $|C_n|^2$.  In Fig.\ 4 the average kinetic energy (a) and force (b) are plotted as a function of time for quadratically contracting box $L=a-bt^2$ at different values of acceleration of the wall $b$. In this case average energy decays as a function of time, while $<F(t)>$ decays. This also can be explained from Fig.\ 5b, where $L(t)$ and $|C_n|^2$ are compared as a function of time.

\begin{figure}[t!]
\centering
\includegraphics[totalheight=0.23\textheight]{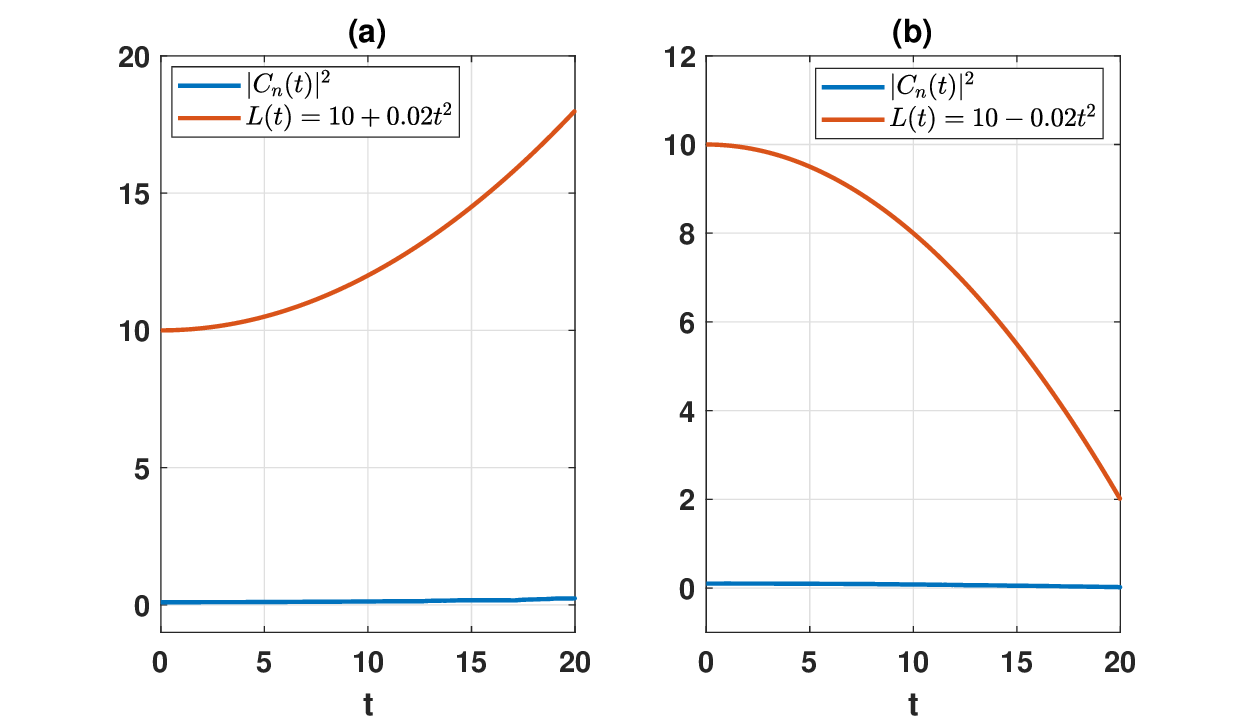}
 \caption{The comparison of time dependence of $|C_n(t)|^2$ and $L(t)$ for expanding box (a) and contracting box (b) for fixed $\alpha=1$} \label{fig5}
\end{figure}

\subsection{Geometric phase}

The case of slowly evolving system in the context of dynamical confinement is of importance. If such evolution occurs due to a slowly varying parameter of the system it can be considered as an adiabatic evolution. An interesting feature of such a system is the existence of so-called geometric phase found by  Berry in the Ref.\ \ci{Berry84}. Berry showed that for a quantum system evolving adiabatically due to a slowly varying parameter, if the parameter  adiabatic changes along the closed curve, $C$ in the parameter space,  the wave function of the system can acquire the so-called geometrical phase in addition to the dynamical one. This additional phase (called later "Berry phase") is different from zero when the trajectory  of the system in the parameter space is located near a point at which the states of the system are degenerate.
 Berry phase for Hermitian counterpart of our model was studied earlier in the Refs.\ \ci{dodonov02,Kwon,Wang}.
For PT-symmetric systems Berry phase becomes complex \ci{Hayward18}-\ci{Fring23}.  In our case, when time-varying parameters is the position of the wall of the box, this corresponds very slowly moving wall, i.e. slowly varying position of the wall, $L(t)$. For slowly moving wall in our model, the Berry phase can be written as \cite{Berry84}
\begin{equation}
    -\frac{1}{2}\frac{\partial^2\Psi_n(L)}{\partial x^2}=E_n(L)\Psi_n(L). \label{beq1}
\end{equation}

\begin{figure}[t!]
\centering
\includegraphics[totalheight=0.2\textheight]{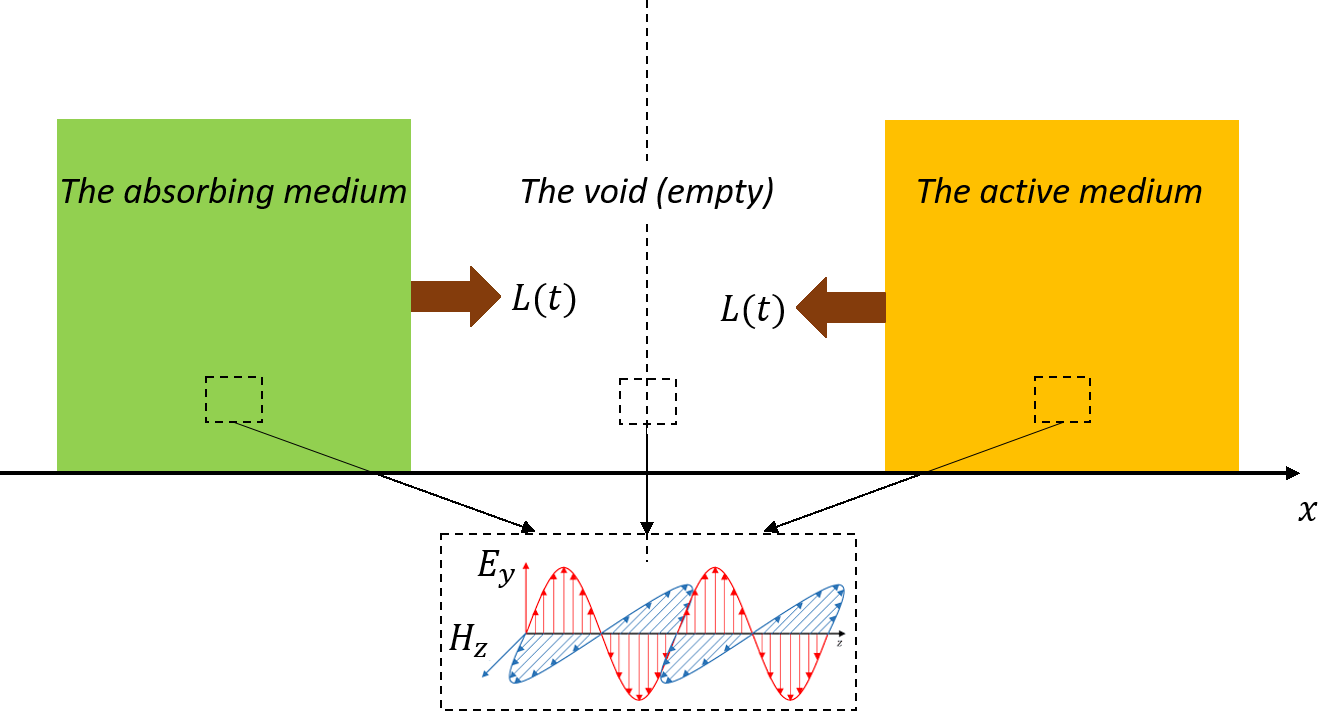}
 \caption{Sketch for the experimental realization of a system ``PT-symmetric particle in time-dependent box".}
 \label{fig6}
\end{figure}

Solution of the Eq.\ (\ref{beq1}) imposed PT-symmetric Robin type boundary conditions (\ref{rbc01}) can be written as
\begin{equation}
    \Psi_n(L)=A_n(L)\left( \sin\frac{\pi n x}{L} +\frac{i\pi n}{L\alpha}\cos\frac{\pi n x}{L} \right) \label{bsol}
\end{equation}
with eigenvalues $E_n(L)=\frac{\pi^2n^2}{2L^2}$ and normalization coefficient $A_n(L)=\sqrt{\frac{L\alpha^2}{L^2\alpha^2+\pi^2n^2}}$.

The geometric phase can be directly computed using its definition  given in \cite{Berry84}:
\begin{equation}
\begin{split}
    & \gamma_n=i\oint\left( \int_{-L(t)}^{L(t)}\Psi_n^*(L)\frac{\partial \Psi_n(L)}{\partial L}dx \right)dL\\
    & =i\int_0^T\left( \int_{-L(t)}^{L(t)}\Psi_n^*(L)\frac{\partial \Psi_n(L)}{\partial L}dx \right) |\dot{L}(t)| dt,
    \end{split}
    \label{bph}
\end{equation}
where $T$ is the period of the function $L(t)$.

In case of the harmonically oscillating wall $L(t)=a+b\cos\omega t$, substituting (\ref{bsol}) into Eq.\ (\ref{bph}) gives us
\begin{equation}
    \gamma_n=i\ln\frac{\alpha^2(a-b)^2+\pi^2n^2}{\alpha^2(a+b)^2+\pi^2n^2}.
\end{equation}
Indeed, $\gamma_n$  is complex for our system. Although Berry phase is computed for a special case of a harmonically oscillating wall, one can do such calculation for different regimes of the wall's motion, provided $L(t)$ for this regime keeps PT-symmetry of the system.

\subsection{A sketch for experimental realization}
Here we discuss ways for experimental realization of our model.
Despite the great progress made in theoretical study of different aspects of PT-symmetric quantum physics, experimental realization of such systems in  quantum mechanical processes has not yet been achieved. 
The only possible way for experimental study of PT-symmetry is using optical systems, who can "mimic" due to the fact that propagation of light in such media can be described in terms of Helmholtz equation formally coinciding with the Schr\"{o}dinger equation. The sketch we propose here also is based on the use of optical media.
Most easiest and direct way could be a sketch presented in Fig.\ 6. It consist of absorbing (loss) and active (gain) materials containing empty space (the void) between them. The motion of the electromagnetic wave, ($E,\;$ $H$) within the void is described in terms of Eq.\ \re{seq1},  with the PT-symmetric boundary condition, given by Eq.\ \re{srbc01}. Moving boundaries can be provided by mechanical way, i.e.\ by shifting the positions of  absorbing and active media. Another option for experimental realization of our model can be using so-called PT-symmetric waveguides \ci{Borisov08,Turitsyna17}, providing  Robin boundary conditions at the edges.
Mechanical motion of the waveguide can produce time-dependent counterpart of such conditions. Finally, the third option could be dynamical version of the so-called electromagnetic analog of the PT-symmetric box, considered in the Ref.\ \ci{Dasarathy13}, where one could consider travelling waves insted of standing ones.

\section{Conclusions}

In this work we studied dynamical confinement in a $PT$-symmetric quantum system created by a quantum box with a moving wall. The system is described in terms of the Schr\"{o}dinger equation with time-varying boundary conditions. The exact solution of the problem is obtained for different regimes of the wall's motion. Such physically observable characteristics as average kinetic energy, average quantum force and average coordinate are computed as a function of time. Geometric 
phase is analytically calculated for an harmonically oscillating wall.
Experimental realization of the model is discussed by proposing a sketch based on the use of PT-symmetric gain and loss media, as well as PT-symmetric waveguides. The model proposed in this paper can be used for the study of different aspects of dynamical confinement in pseudo-Hermitian systems, including Berry phase in different regimes of the wall's motion. The limits of adiabatically moving and sudden removal  of the wall, as well as the case of atom confined in PT-symmetric box are the subjects for forthcoming studies.


\end{document}